\documentclass[Physsubmission, Phys]{SciPost}
\hypersetup{colorlinks,
linkcolor={red!50!black},citecolor={blue!50!black},urlcolor={blue!80!black}}
\usepackage[bitstream-charter]{mathdesign}
\DeclareSymbolFont{usualmathcal}{OMS}{cmsy}{m}{n}
\DeclareSymbolFontAlphabet{\mathcal}{usualmathcal}
\usepackage{fixmath}
\usepackage{graphicx,wrapfig}
\usepackage{amsmath,bm}
\usepackage{slashed}
\allowdisplaybreaks[2]
\usepackage{calrsfs}
\DeclareMathAlphabet{\pazocal}{OMS}{zplm}{m}{n}
\usepackage{color}

\definecolor{mycolor}{rgb}{0.6,0.0,0.4}

\begin{document}

\begin{center}{\Large \textbf{
Transverse-momentum-dependent parton distribution functions\\
for spin-1 hadrons 
}}\end{center}

\begin{center}
S. Kumano\textsuperscript{1,2$\star$} and
Qin-Tao Song\textsuperscript{3} 
\end{center}

\begin{center}
{\bf 1} KEK Theory Center,
             Institute of Particle and Nuclear Studies, KEK,\\
             Oho 1-1, Tsukuba, Ibaraki, 305-0801, Japan
\\
{\bf 2} J-PARC Branch, KEK Theory Center,
             Institute of Particle and Nuclear Studies, KEK, \\
           and Theory Group, Particle and Nuclear Physics Division, 
           J-PARC Center, \\
           Shirakata 203-1, Tokai, Ibaraki, 319-1106, Japan
\\
{\bf 3} School of Physics and Microelectronics, Zhengzhou University, \\
             Zhengzhou, Henan 450001, China
\\
* shunzo.kumano@kek.jp, songqintao@zzu.edu.cn
\end{center}

\begin{center}
\today
\end{center}

\definecolor{palegray}{gray}{0.95}
\begin{center}
\colorbox{palegray}{
  \begin{tabular}{rr}
  \begin{minipage}{0.1\textwidth}
    \includegraphics[width=22mm]{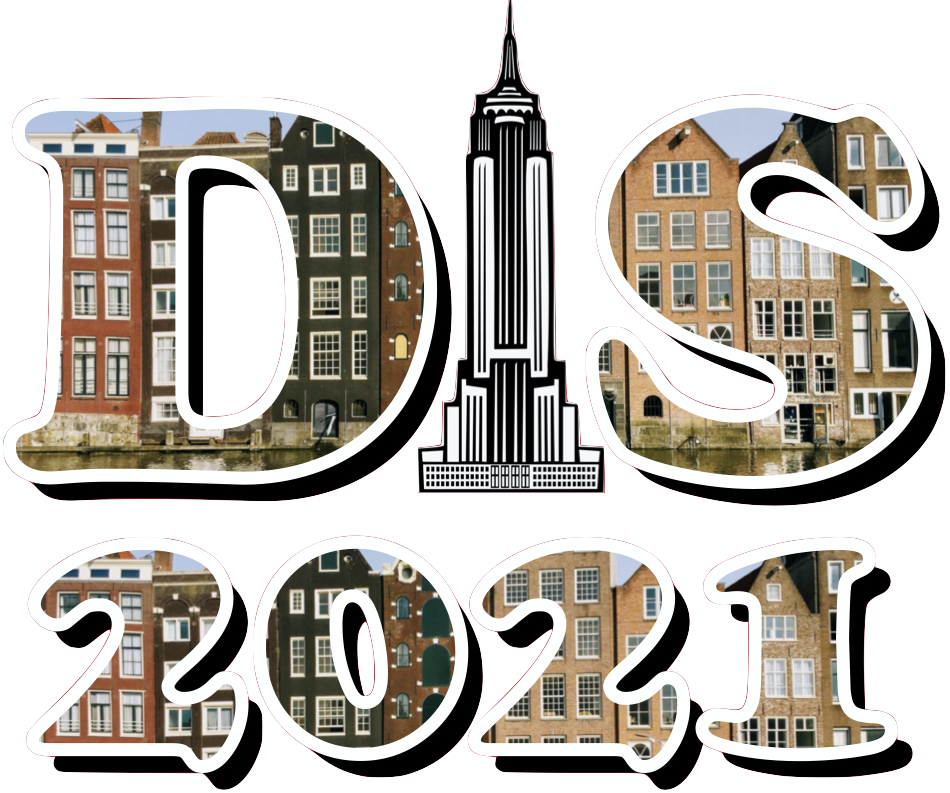}
  \end{minipage}
  &
  \begin{minipage}{0.75\textwidth}
    \begin{center}
    {\it Proceedings for the XXVIII International Workshop\\ on Deep-Inelastic Scattering and
Related Subjects,}\\
    {\it Stony Brook University, New York, USA, 12-16 April 2021} \\
    \doi{10.21468/SciPostPhysProc.?}\\
    \end{center}
  \end{minipage}
\end{tabular}
}
\end{center}

\section*{Abstract}
{\bf
We show transverse-momentum-dependent parton distribution functions (TMDs)
for spin-1 hadrons including twist-3 and 4 functions
by taking the decomposition of a quark correlation function 
in the Lorentz-invariant way with the conditions of Hermiticity 
and parity invariance.
We found 30 new TMDs in the tensor-polarized spin-1 hadron
at twists 3 and 4 in addition to 10 TMDs at twist 2.
Since time-reversal-odd terms of the collinear correlation
function should vanish after integrals over 
the partonic transverse momentum, we obtained new sum rules
for the time-reversal-odd structure functions,
$\bm{\int d^2 k_T g_{LT} 
 = \int d^2 k_T h_{LL} = \int d^2 k_T h_{3LL}}$\\$\bm{=0}$,
at twists 3 and 4.
We also indicated that 
transverse-momentum-dependent fragmentation functions
exist in tensor-polarized spin-1 hadrons.
The TMDs can probe color degrees of freedom, so that 
they are valuable in providing unique opportunities 
for creating interdisciplinary physics fields 
such as gluon condensate, color Aharonov-Bohm effect, 
and color entanglement.
We also found three new collinear PDFs at twists 3 and 4,
and a twist-2 relation and a sum rule were derived
in analogy to the Wandzura-Wilczek relation 
and the Burkhardt-Cottingham sum rule
on the structure function $\bm{g_2}$.
}

\vspace{-0.30cm}
\section{Introduction}
\label{introduction}
\vspace{-0.20cm}

The field of transverse-momentum-dependent parton distribution 
functions (TMDs) is one of rapidly-developing fields in hadron physics. 
It is interesting because the explicit color degrees of freedom 
can be probed by the TMDs. Depending on the color flow 
in hadrons, the TMDs have opposite signs, for example, in semi-inclusive 
deep inelastic scattering and Drell-Yan processes.
The TMD physics is related to fundamental aspects of quantum physics,
such as color Aharonov-Bohm effect and color entanglement,
and the TMDs are also valuable 
for understanding the color glass condensate. 
The color is confined in hadrons and it does not appear easily
in experimental observables. The TMDs provide a unique
opportunity to shed light on new hadron phenomena involving the color.

The TMDs have been investigated so far for the spin-1/2 nucleons.
On the collinear structure function $b_1$ of the spin-1 deuteron,
there was a measurement by the HERMES collaboration in 2005 
\cite{Airapetian:2005cb}.
In future, there are experimental projects to investigate 
structure functions of spin-1 hadrons, especially the spin-1 deuteron,
at the Jefferson laboratory (JLab) \cite{jlab},
Fermilab \cite{Fermilab-dy}, 
and Nuclotron-based Ion Collider fAcility (NICA) \cite{nica},
and electron-ion colliders in US and China.
We also have been investigating structure functions of
spin-1 hadrons theoretically by considering these experimental projects
\cite{b1-sum,our-studies,ks-tmds-2021,ks-twist-2-3-2021}.
Therefore, it is a good opportunity and timely to investigate 
TMDs also for the spin-1 hadrons. However, only the twist-2 
theoretical formalism was available \cite{bm-2000}.
The twist-3 and 4 parts were investigated recently in 
Ref.\,\cite{ks-tmds-2021}, where new structure functions 
and sum rules were proposed at twists 3 and 4.
In addition, a twist-2 relation and a sum rule were derived
for the twist-3 collinear structure function $f_{LT}$ \cite{ks-twist-2-3-2021}.
We explain these works in this paper.

\vspace{-0.35cm}
\section{Correlation functions and polarizations for spin-1 hadrons}
\label{correlation}
\vspace{-0.20cm}

The TMDs and collinear parton distribution functions (PDFs) are
defined from the correlation function given by the matrix element as
\vspace{-0.10cm}
\begin{align}
\Phi_{ij}^{[c]} (k, P, T  \, | \, n )
& =
\int  \! \frac{d^4 \xi}{(2\pi)^4} \, e^{ i k \cdot \xi}
\langle \, P , T \left | \, 
\bar\psi _j (0) \,  W^{[c]} (0, \xi)  
 \psi _i (\xi)  \, \right | P, \,  T \, \rangle .
\label{eqn:correlation-q}
\\[-0.60cm] \nonumber
\end{align} 
Here, $k$ and $P$ are quark and hadron momenta, 
$T$ indicates the tensor polarization of the hadron,
$\xi$ is the space-time coordinate of the quark,
$\psi$ is the quark field, 
$W^{[c]}(0, \xi)$ is the gauge link,
$c$ indicates the integral path,
and $n$ is the lighcone vector
$n^\mu =(1,0,0,-1)/\sqrt{2}$.
Since only the tensor polarization is considered in this work,
the vector polarization $S$ is not explicitly written.
The TMD and collinear correlation functions are defined
by integrating the quark transverse momenta as
\vspace{-0.10cm}
\begin{align}
\Phi^{[c]} (x, k_T, P, T ) & = \! \int \! dk^+ dk^- \, 
               \Phi^{[c]} (k, P, T  \, |n ) \, \delta (k^+ \! -x P^+) ,
\label{eqn:correlation-tmd}
\\[-0.05cm]
\Phi (x, P, T ) & 
  = \! \int \! d^2 k_T \, \Phi^{[c]} (x, k_T, P, T ) .
\label{eqn:correlation-pdf}
\\[-0.70cm] \nonumber
\end{align}

The covariant form of the tensor polarization $T^{\mu\nu}$ 
of a spin-1 hadron is expressed 
by the tensor-polarization parameters $S_{T}^x$, $S_{T}^y$, $S_L$, 
$S_{LL}$, $S_{TT}^{xx}$, $S_{TT}^{xy}$, $S_{LT}^x$, and $S_{LT}^y$
as 
\vspace{-0.05cm}
\begin{align}
T^{\mu\nu}  = \frac{1}{2} &  \left [ \frac{4}{3} S_{LL} \frac{(P^+)^2}{M^2} 
               \bar n^\mu \bar n^\nu 
          - \frac{2}{3} S_{LL} ( \bar n^{\{ \mu} n^{\nu \}} -g_T^{\mu\nu} )
\right.
\nonumber \\[-0.10cm]
& 
\left.
+ \frac{1}{3} S_{LL} \frac{M^2}{(P^+)^2}n^\mu n^\nu
+ \frac{P^+}{M} \bar n^{\{ \mu} S_{LT}^{\nu \}}
- \frac{M}{2 P^+} n^{\{ \mu} S_{LT}^{\nu \}}
+ S_{TT}^{\mu\nu} \right ].
\label{eqn:spin-1-tensor-1}
\\[-0.70cm] \nonumber
\end{align}
Here, $a^{\{ \mu} b^{\nu \}}$ indicates the symmetrized combination
$a^{\{ \mu} b^{\nu \}} = a^\mu b^\nu + a^\nu b^\mu$,
$M$ is the hadron mass, and $P^+$ is the lightcone momentum
given by $P^+ =(P^0 +P^3)/\sqrt{2}$.
Using this tensor together with available Lorentz vectors $P$, $k$, and $n$, 
we wrote the general form of the correlation function
by considering the condition of Hermiticity and parity invariance as
\cite{ks-tmds-2021}
\begin{align}
\Phi(k, P, T \, | n) & = \frac{A_{13}}{M}  T_{kk} + \frac{A_{14}}{M^2} T_{kk} 
     \slashed{P}
+ \cdots
+ \frac{A_{20}}{M^2} \varepsilon^{\mu\nu P k}  \gamma_{\mu} \gamma_5 T_{\nu k}
\nonumber \\
&
+ \frac{B_{21}M}{P\cdot n} T_{kn}  +\frac{B_{22}M^3}{(P\cdot n)^2} T_{nn}
+ \cdots
+ \frac{B_{52}M}{P\cdot n } \sigma_{\mu k}  T^{\mu n} ,
\label{eqn:cork4}
\end{align} 
where the notation $X_{\mu k} \equiv X_{\mu \nu} k^{\nu}$ 
is used for the contraction of the tensor $X_{\mu \nu}$ with $k^{\nu}$. 
The important point of this expression is that the terms
with the lightcone vector $n$ are included for finding
the twist-3 and 4 structure functions, as investigated
in finding twist-3 and 4 structure functions
in the spin-1/2 nucleons \cite{tmds-nucleon}.
Integrating this expression as given in Eq.\,(\ref{eqn:correlation-tmd}),
we obtained possible tensor-polarized structure functions up to twist 4.

\section{TMDs and PDFs of spin-1 hadrons up to twist 4}
\label{TMDs-PDFs}
\vspace{-0.20cm}

\noindent
{\bf\large TMDs for tensor-polarized spin-1 hadrons}
\vspace{0.10cm}

\begin{wraptable}[25]{r}{0.42\textwidth}
   \vspace{-1.3cm}
   \hspace{-0.0cm}
\begin{center}
  \includegraphics[width=6.4cm]{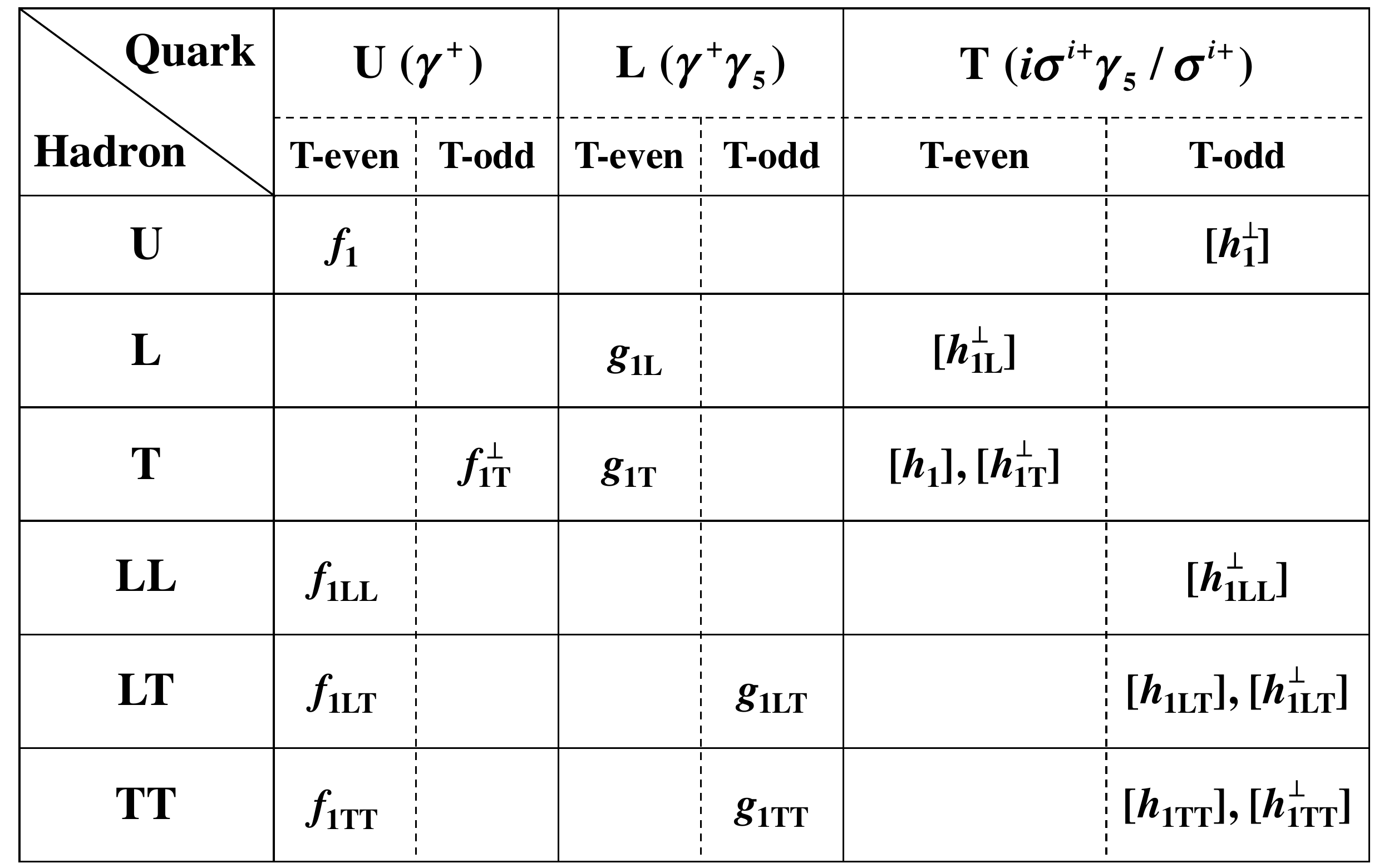}
\end{center}
\vspace{-0.50cm}
\caption{\hspace{-0.10cm} Twist-2 TMDs.}
\label{table:twist-2-tmds}
\ \\[-0.20cm]
  \includegraphics[width=6.4cm]{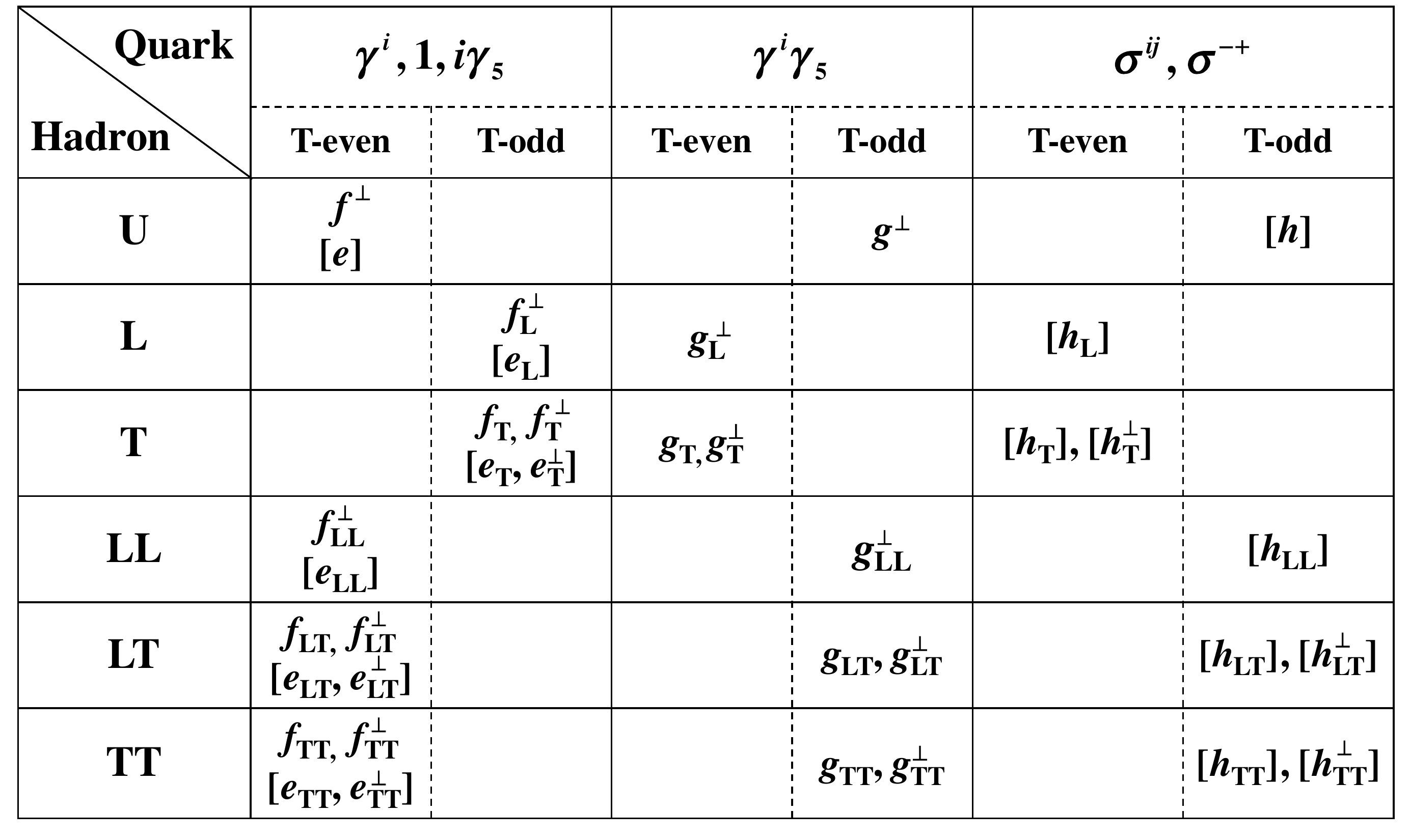}
\vspace{-0.30cm}
\caption{\hspace{-0.10cm} Twist-3 TMDs.}
\label{table:twist-3-tmds}
\ \\[-0.20cm]
  \includegraphics[width=6.4cm]{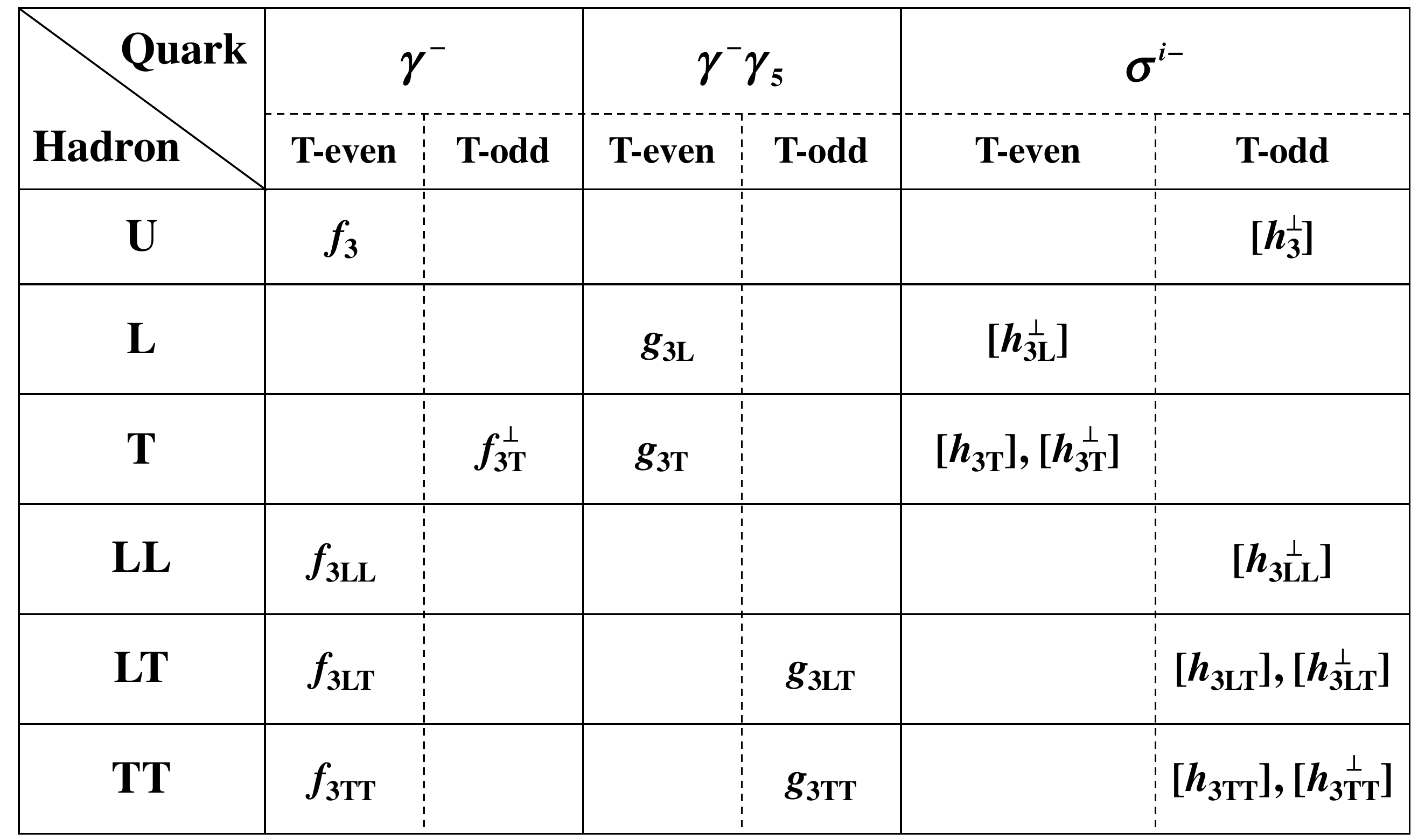}
\vspace{-0.30cm}
\caption{\hspace{-0.10cm} Twist-4 TMDs.}
\label{table:twist-4-tmds}
\end{wraptable}

The TMDs are defined from the correlation functions
of Eqs.\,(\ref{eqn:correlation-tmd}) and (\ref{eqn:cork4}) 
by taking traces with various $\gamma$ matrices as 
$
\Phi^{\left[ \Gamma \right]} (x, k_T, T) \equiv 
\frac{1}{2} \, \text{Tr} \left[ \, 
\Phi(x, k_T, T ) 
\Gamma \, \right] ,
$
and possible TMDs are listed in Tables \ref{table:twist-2-tmds},
\ref{table:twist-3-tmds}, and \ref{table:twist-4-tmds}.
The upper half sections (unpolarized U, longitudinally polarized L,
transversely polarized T) of these tables are the same as
the ones of the spin-1/2 nucleons.
In this work, the tensor-polarized sections (LL, LT, TT) are studied.
The square brackets $[\ ]$ indicate chiral-odd distributions
and the others are chiral-even ones. 
The T-even (T: time reversal) and T-odd distributions are separately written.
The twist-2 part was studied in Ref.\,\cite{bm-2000}
by calculating $\Phi^{ [ \gamma^+ ] }$,
$\Phi^{ [ \gamma^+ \gamma_5 ] }$, and
$\Phi^{ [ i \sigma^{i+} \gamma_5 ] }$
($\Phi^{ [ \sigma^{i+} ] }$).
Our new results were for the twist-3 and twist-4 functions
in Tables \ref{table:twist-3-tmds} and \ref{table:twist-4-tmds}
\cite{ks-tmds-2021}.

Since the full expressions are lengthy, we explain the outline
by taking an example on the correlation function $\Phi^{ [ \gamma^i ] }$
and related twist-3 TMDs. The twist-3 TMDs were obtained by calculating
$\Phi^{ [ \gamma^i ] }$,
$\Phi^{\left[{\bf 1}\right]}$,
$\Phi^{\left[i\gamma_5\right]}$
$\Phi^{ [\gamma^{i}\gamma_5 ]}$
$\Phi^{ [ \sigma^{ij} ]}$,
and $\Phi^{ [ \sigma^{-+} ] }$.
Among them, $\Phi^{ [ \gamma^i ] }$ is expressed 
by the polarization parameters and TMDs as
\vspace{-0.60cm}
\begin{align}
& 
\Phi^{ [ \gamma^i ] } (x, k_T, T)
= 
\frac{M}{P^+} \bigg [  f^{\perp}_{LL}(x, k_T^{\, 2})  S_{LL} \frac{k_T^i}{M}
\! + \! f^{\,\prime} _{LT} (x, k_T^{\, 2})S_{LT}^i 
\nonumber \\[-0.10cm]
& \ \hspace{1.0cm}
- f_{LT}^{\perp}(x, k_T^{\, 2}) \frac{ k_{T}^i  S_{LT}\cdot k_{T}}{M^2} 
- f_{TT}^{\,\prime} (x, k_T^{\, 2}) \frac{S_{TT}^{ i j} k_{T \, j} }{M} 
\nonumber \\[-0.10cm]
& \ \hspace{1.0cm}
+ f_{TT}^{\perp}(x, k_T^{\, 2}) \frac{k_T\cdot S_{TT}\cdot k_T}{M^2} 
       \frac{k_T^i}{M} \bigg ] ,
\label{eqn:cork-3-1a}
\nonumber \\[-0.70cm]
\end{align} 
where the TMDs $f^{\perp}_{LL}$, $\cdots$, $f_{TT}^{\perp}$
are expressed by the expansion coefficients
$A_i$ and $B_i$ of Eq.\,(\ref{eqn:cork4}).
The guideline for assigning the TMD names 
$f^{\perp}_{LL}$, $\cdots$, $f_{TT}^{\perp}$
is explained in Ref.\,\cite{ks-tmds-2021}
depending on polarizations and transverse-momentum factors.
In this way, we obtained five TMDs in $\Phi^{ [ \gamma^i ] }$
and they are listed in Table \ref{table:twist-3-tmds}.
In listing the TMDs in the tables, we redefined the TMDs by 
$
F (x, k_T^{\, 2}) \equiv F^{\,\prime} (x, k_T^{\, 2})
 - (k_T^{\, 2} /(2M^2)) \, F^{\perp} (x, k^{\, 2}_T) 
$
where $k_T^{\, 2}= - \vec k_T^{\, 2}$,
and the functions without $^\prime$ are shown in the tables.
The other twist-3 TMDs are also listed in Table \ref{table:twist-3-tmds}.
In the same way, the twist-4 TMDs are obtained by calculating
$\Phi^{[\gamma^-]}$, $\Phi^{[\gamma^- \gamma_5]}$, and $\Phi^{[\sigma^{i-}]}$
and they are listed in Table \ref{table:twist-4-tmds}.
There are 40 TMDs in the tensor-polarized spin-1 hadron
as shown in Tables \ref{table:twist-2-tmds}, \ref{table:twist-3-tmds},
and \ref{table:twist-4-tmds}. Among them, we found that
30 new TMDs exist at twist-3 and 4 as
\vspace{-0.10cm}
\begin{align}
& \text{Twist-3 TMD:}\ \ f_{LL}^\perp,\ e_{LL},\  
      f_{LT},\ f_{LT}^\perp,\ e_{1T},\ e_{1T}^\perp,\ 
      f_{TT},\ f_{TT}^\perp,\ e_{TT},\ e_{TT}^\perp,\ 
      g_{LL}^\perp,\ g_{LT},\ g_{LT}^\perp,
\nonumber \\
& \ \hspace{2.3cm}
      g_{TT},\ g_{TT}^\perp,\ 
      h_{1L},\ h_{LT},\ h_{LT}^\perp,\ h_{TT},\ h_{TT}^\perp,
\nonumber \\
& \text{Twist-4 TMD:}\ \ f_{3LL},\ f_{3LT},\ f_{3TT},\ g_{3LT},\ g_{3TT},\  
      h_{3LL}^\perp,\ h_{3LT},\ h_{3LT}^\perp,\ h_{3TT},\ h_{3TT}^\perp .
\label{eqn:spin-1-tmds-3-4}
\end{align} 
They are classified by the time-reversal and chirality properties.
Since the T-odd collinear PDFs should vanish 
$f (x)_{\text{T-odd}}=0$
due to the time-reversal invariance, we have the following sum rules
at twist-3 and 4
\vspace{-0.20cm}
\begin{align}
\! \int \! d^2 k_T \, g_{LT} (x, k_T^{\, 2}) 
= \! \int \! d^2 k_T \, h_{LL} (x, k_T^{\, 2}) 
= \! \int \! d^2 k_T \, h_{3LT}(x, k_T^{\, 2})  = 0 .
\label{eqn:TMD-sum}
\end{align} 

\vspace{0.00cm}
\noindent
{\bf\large TMD fragmentation functions of spin-1 hadrons}
\vspace{0.10cm}

New TMD fragmentation functions also exist for the spin-1 hadrons
\cite{ks-tmds-2021},
and they are obtained simply by changing the kinematical variables
and function names as
\vspace{-0.20cm}
\begin{align}
& \ \hspace{-0.20cm}
\text{Kinematical variables:}   \ \  
x, k_T, S, T, M, n, \gamma^+, \sigma^{i+}
\Rightarrow \ 
 z, k_T, S_h, T_h, M_h, \bar n, \gamma^-, \sigma^{i-},
\nonumber \\
& \ \hspace{-0.20cm}
\text{Distribution functions:}  \ \ f, g, h, e \hspace{2.30cm}
\ \Rightarrow \ 
\text{Fragmentation functions:} \ 
D, G, H, E .
\label{eqn:tmd-fragmentation}
\end{align} 
\vspace{-0.70cm}

\begin{wraptable}[29]{r}{0.38\textwidth}
   \vspace{-0.7cm}
   \hspace{-0.0cm}
\begin{center}
  \includegraphics[width=5.8cm]{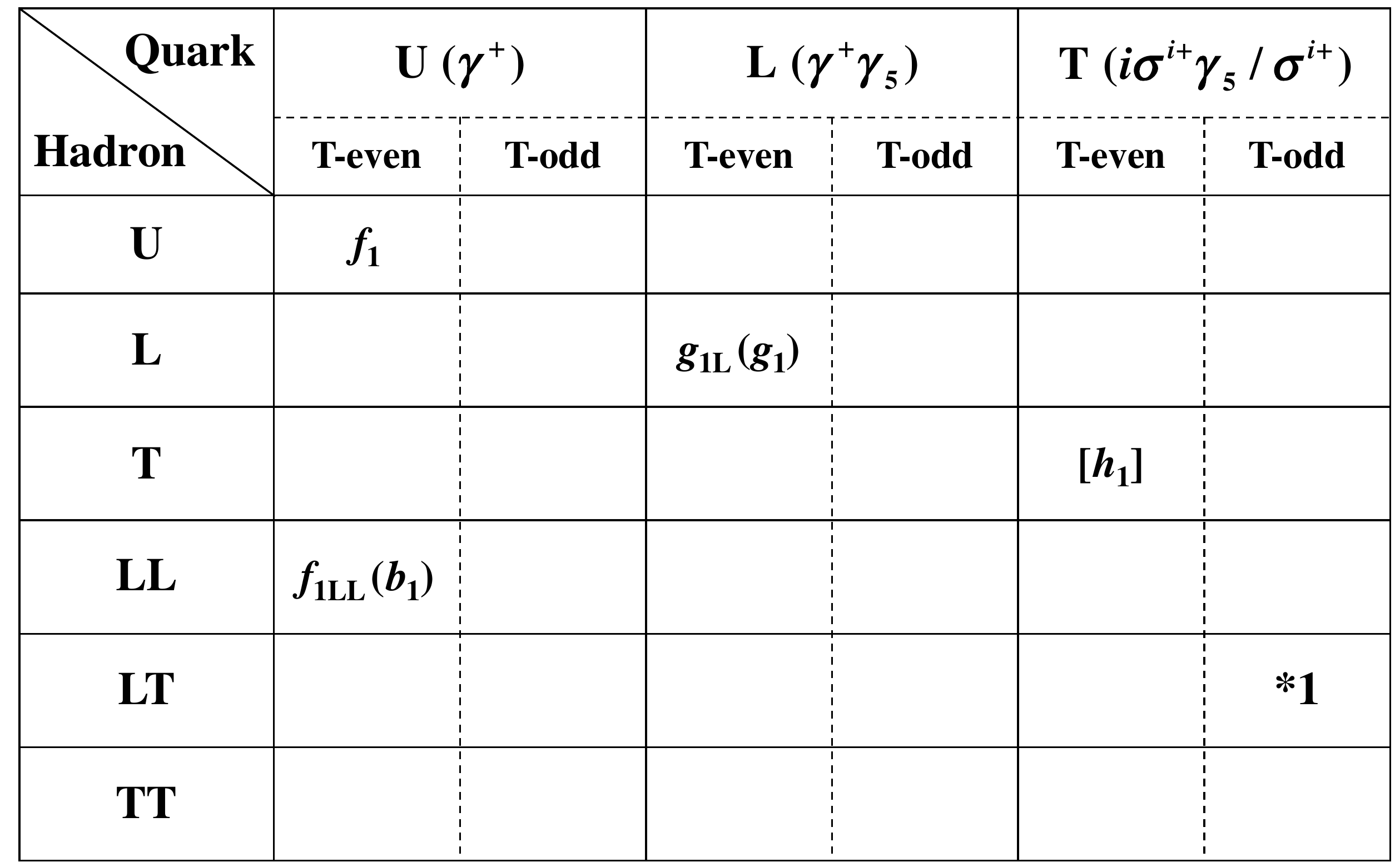}
\end{center}
\vspace{-0.50cm}
\caption{\hspace{-0.10cm} Twist-2 PDFs.}
\label{table:twist-2-pdfs}
\ \\[-0.20cm]
  \includegraphics[width=5.8cm]{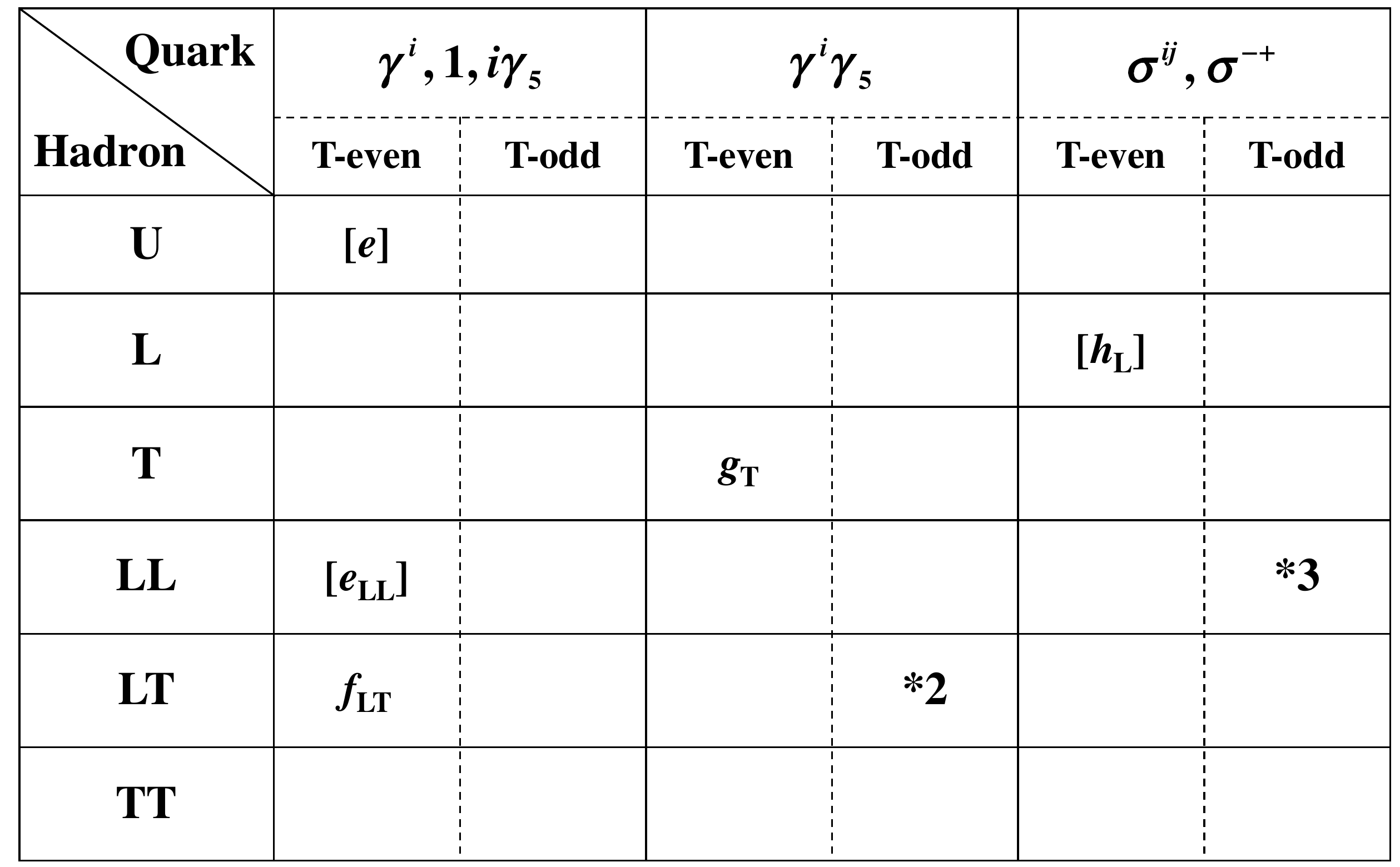}
\vspace{-0.30cm}
\caption{\hspace{-0.10cm} Twist-3 PDFs.}
\label{table:twist-3-pdfs}
\ \\[-0.20cm]
  \includegraphics[width=5.8cm]{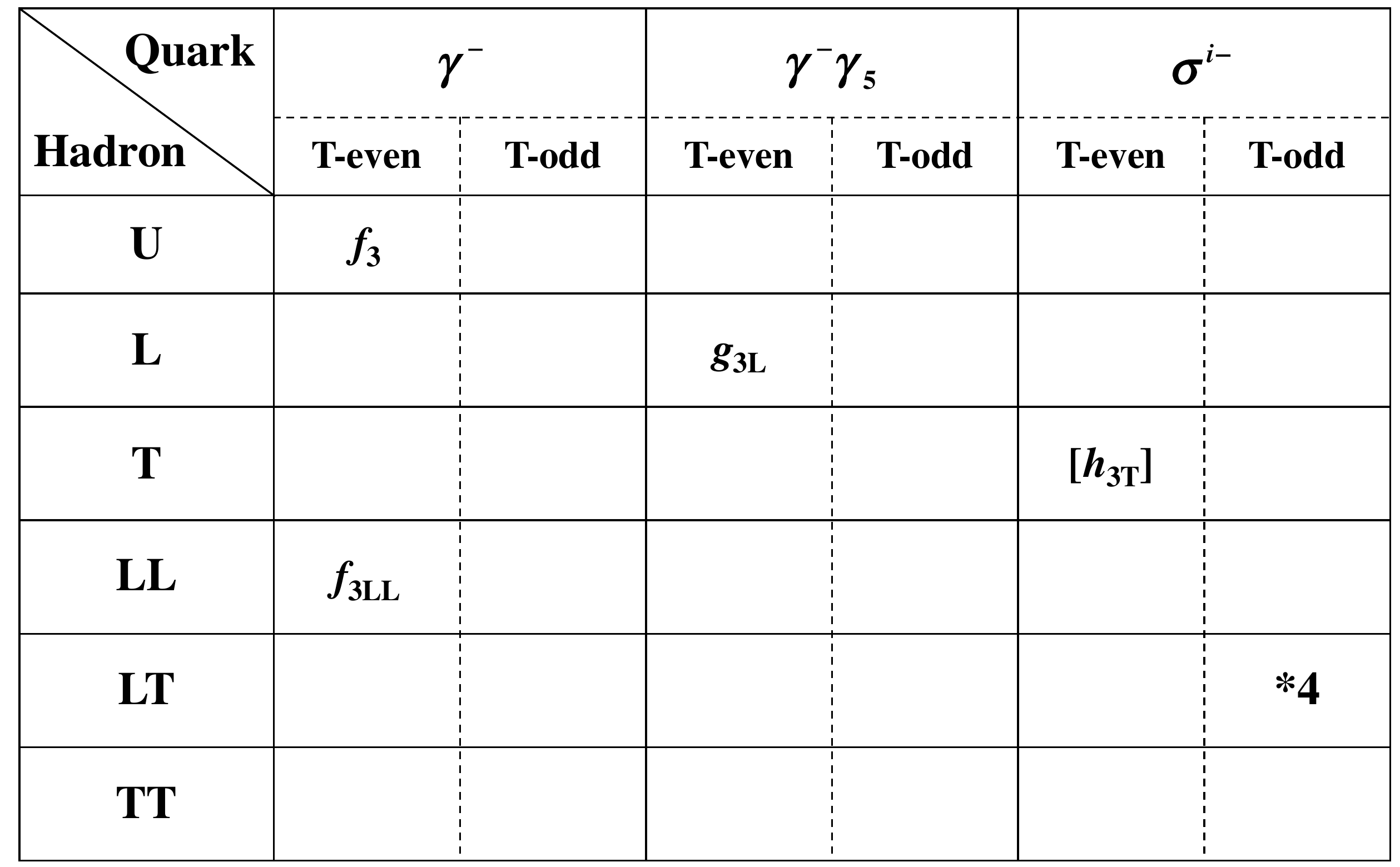}
\vspace{-0.30cm}
\caption{\hspace{-0.10cm} Twist-4 PDFs.}
\label{table:twist-4-pdfs}
\end{wraptable}

\vspace{0.20cm}
\noindent
{\bf\large PDFs for tensor-polarized spin-1 hadrons}
\vspace{0.10cm}

The collinear PDFs of the tensor-polarized hadrons 
\cite{deuteron-tensor}
are obtained
by integrating the TMDs over the transverse momentum as
$f (x) = \int d^2 k_T f (x, k_T^{\, 2})$.
Many functions vanish by this integral and the remaining
PDFs are shown in Tables \ref{table:twist-2-pdfs},
\ref{table:twist-3-pdfs}, and \ref{table:twist-4-pdfs}.
Therefore, the new twist-3 and 4 functions, which we found, are 
\cite{ks-tmds-2021}
\vspace{-0.20cm}
\begin{align}
\text{Twist-3 PDF:}\ e_{LL},\ f_{LT},\ \ \  
\text{Twist-4 PDF:}\ f_{3LL}. 
\label{eqn:pdfs-3-4}
\end{align} 
\vspace{-0.70cm}

\noindent
The asterisks $*1$,\,$*2$,\,$*3$,\,$*4$ indicate
the collinear PDFs
$h_{1LT} (x)$,\,$g_{LT} (x)$,\,$h_{LL} (x)$,\,$h_{3LT} (x)$
vanish, respectively, because of the time-reversal invariance.
However, since the time-reversal invariance cannot be imposed
in the fragmentation functions, the corresponding fragmentation 
functions
$H_{1LT} (z)$,\,$G_{LT} (z)$,\,$H_{LL} (z)$,\,$H_{3LT} (z)$, 
as indicated by the replacements of 
Eq.\,(\ref{eqn:tmd-fragmentation}), 
should exist as collinear fragmentation functions
\cite{ks-tmds-2021,ji-ffs}.

\vspace{0.20cm}
\noindent
{\bf Twist-2 relation and sum rule for $\bm f_{LT}$}
\vspace{0.10cm}

Since we obtained the new PDFs for spin-1 hadrons, it is possible 
to investigate useful relations for them in the similar way
to the Wandzura-Wilczek (WW) relation 
and the Burkhardt-Cottingham (BC) sum rule.
For finding twist-2 relations, it is important to
specify higher-twist effects. Twist-3 effects are
investigated by the nonlocal operator 
$\bar{\psi}(0) \big (
    \partial^{\mu} \gamma^{\alpha} 
   -\partial^{\alpha} \gamma^{\mu} \big ) \psi (\xi)$,
whose matrix element between tensor-polarized spin-1 hadron states
is expressed by $f_{1LL}$ and $f_{LT}$ up to twist 3.
For the tensor-polarized spin-1 hadron,
twist-3 multiparton distribution functions 
\vspace{-0.20cm}
\begin{align}
F_{LT} (x,y),\ \ G_{LT} (x,y),\ \  
H_{LL}^\perp (x,y),\ \  H_{TT} (x,y) ,
\end{align}
\vspace{-0.70cm}

\noindent
exist in general \cite{ks-twist-2-3-2021}.
The twist-3 matrix element is expressed by $F_{G,LT}$ and $G_{G,LT}$
defined with the gluon field tensor $G^{\mu\nu}$.
Specifying the twist-3 terms, we obtained
\cite{ks-twist-2-3-2021}
\vspace{-0.20cm}
\begin{align}
f_{LT}(x) & = \frac{3}{2} \int^{\epsilon (x)}_x dy \frac{f_{1LL}(y)}{y}
          +\int^{\epsilon (x)}_x dy \frac{f_{LT}^{(HT)}(y)}{y},
\label{eqn:fbar-twist-3}
\end{align}
\vspace{-0.30cm}

\noindent
where the higher-twist function $f_{LT}^{(HT)} (x)$ is expressed 
by the integral of $F_{G,LT}$ and $G_{G,LT}$.
Here, $\epsilon (x)=1 \ (-1)$ for $x>0 \ (x<0)$.
In the positive (negative) $x$ region, $f(x)$ is a quark (antiquark)
distribution.
We define $f^+$ distribution by
$ f^+ (x) \equiv f (x) + \bar f (x) =  f (x) -  f (-x) $
for 
$f=f_{1LL},\ f_{LT},\ f_{LT}^{(HT)}$ at $x>0$, 
the relation becomes
$
f_{LT}^+(x)= \frac{3}{2} \int^1_x dy \, f_{1LL}^+(y)/y 
$
if the higher-twist terms are neglected.
Furthermore, if the function $f_{2LT}$ is defined by
$f_{2LT}(x) \equiv  (2/3) f_{LT}(x) \\ - f_{1LL}(x)$,
this integral relation becomes
\vspace{-0.20cm}
\begin{align}
f_{2LT}^+ (x)=-f_{1LL}^+ (x)
              + \int^1_x \frac{dy}{y} f_{1LL}^+ (y) .
\label{eqn:f2lt-relation}
\end{align}
\vspace{-0.25cm}

\noindent
We obtain a sum rule by integrating this twist-2 relation over $x$ as 
\vspace{-0.15cm}
\begin{align}
\int_0^1 dx \, f_{2LT}^+(x) =0 .
\end{align}
\vspace{-0.25cm}

\noindent
These are analogous equations to the WW relation and BC sum rule.
The function $f_{1LL}$ is given by the tensor-polarized structure 
function $b_1$ as $-(3/2) f_{1LL}^+ = b_1^q + b_1^{\bar q}$.
If the tensor-polarized antiquark distributions vanish,
the parton-model sum rule $\int dx b_1 =0$
exists \cite{b1-sum}. Then, there is a sum rule 
for the $f_{LT}$ itself as
$ \int_0^1 dx \, f_{LT}^+(x) =0$.

\vspace{-0.45cm}
\section{Conclusion}
\label{conclusion}
\vspace{-0.30cm}

We proposed new TMD structure functions at twist 3 and 4
by decomposing the TMD correlation function in the Lorentz invariant way
with constraints of Hermiticity and parity invariance.
There are 40 TMDs for the tensor-polarized spin-1 hadron up to twist 4,
and we found new 30 TMDs at twist 3 and 4. 
Then, we showed sum rules for the time-reversal-odd TMDs.
There exist also corrensponing fragmentation functions 
simply changing kinematical variables and function names.
Integrating the TMDs over the transverse momentum, we obtained
new collinear PDFs $e_{LL}$, $f_{LT}$, and $f_{3LL}$ at twist 3 and 4,
in addition to the twist-2 function $f_{1LL}$ ($b_1$).
We also indicated the twist-2 relation and the sum rule for $f_{LT}$
in analogy to the Wandzura-Wilczek relation 
and the Burkhardt-Cottingham sum rule.

\vspace{-0.45cm}
\section*{Acknowledgements}
\vspace{-0.30cm}

S. Kumano was partially supported by 
Japan Society for the Promotion of Science (JSPS) Grants-in-Aid 
for Scientific Research (KAKENHI) Grant Number 19K03830.
Qin-Tao Song was supported by the National Natural Science Foundation 
of China under Grant Number 12005191 and the Academic Improvement Project 
of Zhengzhou University.

\vspace{-0.45cm}



\begin{thebibliography}{99}
\vspace{-0.30cm}
\bibitem{Airapetian:2005cb} 
    A.~Airapetian \textit{et al.} (HERMES Collaboration),
          Phys. Rev. Lett. {\bf 95} (2005) 242001.
\vspace{-0.15cm}
\bibitem{jlab} 
J.-P. Chen {\it et al.},
       {Proposal to Jefferson Lab PAC-38, PR12-11-110 (2011)};
    M. Jones {\it et al.},
    {A Letter of Intent to Jefferson Lab PAC 44, LOI12-16-006 (2016)}.
\vspace{-0.15cm}
\bibitem{Fermilab-dy} 
    D. Geesaman {\it et al.},
    {Proposal to Fermilab PAC, P-1039 (2013)}.
\vspace{-0.15cm}
\bibitem{nica}
    A. Arbuzov {\it et al.},
    Prog. Nucl. Part. Phys. {\bf 119}, 103858 (2021) 1.
\vspace{-0.15cm}
\bibitem{b1-sum}
F. E. Close and S. Kumano, 
      Phys. Rev. D {\bf 42} (1990) 2377.
\vspace{-0.15cm}
\bibitem{our-studies}
S. Hino and S. Kumano, 
    Phys. Rev. D {\bf 59} (1999) 094026;
    D {\bf 60} (1999) 054018;
T.-Y. Kimura and S. Kumano,
    Phys. Rev. D {\bf 78} (2008) 117505;
S. Kumano, 
    Phys. Rev. D {\bf 82} (2010) 017501;
W. Cosyn, Yu-Bing Dong, S. Kumano, and M. Sargsian,
       Phys. Rev. D {\bf 95} (2017) 074036;
S.~Kumano and Qin-Tao Song,
    Phys. Rev. D {\bf 94} (2016) 054022;
     D {\bf 101} (2020) 054011 \& 094013.
\vspace{-0.15cm}
\bibitem{ks-tmds-2021}
S.~Kumano and Qin-Tao Song,
    Phys. Rev. D {\bf 103} (2021) 014025.
\vspace{-0.15cm}
\bibitem{ks-twist-2-3-2021}
S.~Kumano and Qin-Tao Song,
    {arXiv:2106.15849}, submitted for publication.
\vspace{-0.15cm}
\bibitem{bm-2000} 
A. Bacchetta and P. Mulders 
     Phys. Rev. D {\bf 62} (2000) 114004.
\vspace{-0.15cm}
\bibitem{tmds-nucleon}
  K.~Goeke, A.~Metz and M.~Schlegel,
      Phys. Lett. B {\bf 618} (2005) 90;
  A.~Metz, P.~Schweitzer and T.~Teckentrup,
      Phys. Lett. B {\bf 680} (2009) 141.
\vspace{-0.15cm}
\bibitem{deuteron-tensor}
  P.~Hoodbhoy, R.~L.~Jaffe, and A.~Manohar,
      Nucl. Phys. B {\bf 312} (1989) 571;
  J.~P.~Ma, C.~Wang, and G.~P.~Zhang, 
      arXiv:1306.6693 (2013).
\vspace{-0.15cm}
\bibitem{ji-ffs} 
  X. Ji, Phys. Rev. D {\bf 49} (1994) 114.
\end{thebibliography}
\end{document}